\documentclass[iop]{emulateapj-rtx4}
\usepackage{graphicx}
\usepackage{amssymb}
\usepackage{multirow}
\usepackage{float}
\usepackage{comment}
\usepackage{booktabs}
\usepackage[normalem]{ulem}
\usepackage{bm}

\newcommand{\be}{\begin{equation}}
\newcommand{\ee}{\end{equation}}

\begin{document}
 
\title{{\sl XMM-Newton} Observations of Young and Energetic Pulsar J2022+3842}

\author{P. Arumugasamy}
\affil{Department of Astronomy \& Astrophysics, Pennsylvania State University, 525 Davey Lab,University Park, PA 16802, USA }
\email{prakash@astro.psu.edu}
\author{G. G. Pavlov}
\affil{Department of Astronomy \& Astrophysics, Pennsylvania State University, 525 Davey Lab,University Park, PA 16802, USA}
\author{O. Kargaltsev}
\affil{Department of Physics, The George Washington University, Washington, DC 20052, USA}
 
\begin{abstract}
We report on {\sl XMM-Newton} EPIC observations of the young pulsar J2022+3842, with a characteristic age of 8.9 kyr.
We detected X-ray pulsations and found the pulsation period $P\approx 48.6$ ms, and its derivative $\dot{P}\approx 8.6\times 10^{-14}$, twice larger than the previously reported values.
The pulsar exhibits two very narrow (FWHM $\sim 1.2$ ms) X-ray pulses each rotation, separated by $\approx 0.48$ of the period, with a pulsed fraction of $\approx 0.8$.
Using the correct values of $P$ and $\dot{P}$, we calculate the pulsar's spin-down power $\dot{E}=3.0 \times 10^{37}$ erg s$^{-1}$ and magnetic field $B=2.1\times 10^{12}$ G.
The pulsar spectrum is well modeled with a hard power-law (PL) model (photon index $\Gamma = 0.9\pm0.1 $, hydrogen column density $n_H = (2.3\pm0.3) \times 10^{22}\,{\rm cm}^{-2}$).
We detect a weak off-pulse emission which can be modeled with a softer PL ($\Gamma \approx 1.7\pm0.7$), poorly constrained because of contamination in the EPIC-pn timing mode data. 
The pulsar's X-ray efficiency in the 0.5--8 keV energy band, $\eta_{\rm PSR}= L_{\rm PSR}/\dot{E} = 2 \times 10^{-4} (D/10\,{\rm kpc})^2$, is similar to those of other pulsars.
The {\sl XMM-Newton} observation did not detect extended emission around the pulsar.
Our re-analysis of {\sl Chandra} X-ray observatory archival data shows a hard, $\Gamma \approx 0.9 \pm 0.5$, spectrum and a low efficiency, $\eta_{\rm PWN}\sim 2\times 10^{-5} (D/10\,{\rm kpc})^2$, for the compact pulsar wind nebula, unresolved in the {\sl XMM-Newton} images.

\end{abstract}


\section{Introduction}
\setcounter{footnote}{0}
Nonthermal emission of rotation-powered pulsars (RPPs), observable from the radio to $\gamma$-rays, is powered by the loss of their rotational energy. 
X-ray observations of RPPs allow one to understand the origin and mechanisms of the nonthermal emission from the pulsar magnetosphere and thermal emission from the neutron star (NS) surface. 
If the pulsar is young enough, X-ray observations can also detect the pulsar wind nebula (PWN), whose synchrotron emission is generated by relativistic particles outflowing from the pulsar magnetosphere, and the supernova remnant (SNR), formed by the same supernova explosion as the pulsar.
They are particularly useful for pulsars that have been observed at other wavelengths, in which case the multi-wavelength data analysis helps to understand the properties of the emitting particles, the locations of the emitting regions, and the mechanisms involved in the multi-wavelength emission.

PSR J2022+3842 is a young, energetic pulsar, discovered by \cite{2011ApJ...739...39A} (henceforth referred to as A+11) in a 54 ks {\sl Chandra} X-ray observatory (CXO) observation of the radio SNR G76.9+1.0 \citep{1993A&A...276..522L}.
Although A+11 found no evidence for G76.9+1.0 in the CXO data, they did find a point source CXOU\,J202221.68+384214.8, surrounded by a faint nebulosity, at the center of the radio SNR, which they interpreted as a pulsar and its PWN.
A+11 fit an absorbed power-law (PL) model to the pulsar spectrum and found a hydrogen column density $n_{H,22}\equiv n_H/(10^{22}\,{\rm cm}^{-2}) = 1.6 \pm 0.3$ and a photon index $\Gamma = 1.0 \pm 0.2$. 
From an absorbed PL fit of the PWN spectrum, they obtained an unusually low absorbed flux ratio $F_{\rm PWN}/F_{\rm PSR} \approx 0.08$ in the 2--10 keV band (assuming fixed $n_{H,22}= 1.6$ and $\Gamma = 1.4$ parameter values).

From follow-up observations in the radio with the Green Bank Telescope (GBT) and in X-rays with the {\sl Rossi X-ray Timing Explorer} ({\sl RXTE}), A+11 found a pulsation period $P=24$ ms with a spin-down rate $\dot{P}\approx 4.3 \times 10^{-14}\;{\rm s\;s}^{-1}$ (MJD 54957--55469), and a spin glitch of magnitude $\Delta P/P \simeq 1.9 \times 10^{-6}$ (between MJD 54400 and 54957).
They derived the pulsar's dispersion measure DM $= 429.1 \pm 0.5 \;{\rm pc\; cm}^{-3}$, which formally corresponds to very large distances, $D>50$ kpc in the NE2001 Galactic electron distribution model \citep{2002astro.ph..7156C}.
However, the authors noted that a likely overdensity of free electrons in the Cygnus region, along the line of site, may account for the higher-than-expected DM, so the actual distance remains uncertain. 

The pulsar's 2--20 keV X-ray pulse profile, obtained with the GBT/{\sl RXTE} ephemeris, shows a single narrow pulse (FWHM = 0.06 of full cycle) with a 91\% -- 100\% pulsed fraction (A+11).
The authors fit a $\Gamma = 1.1 \pm 0.2$ PL model to the pulsed spectrum with fixed $n_{H,22}=1.6$.
They derived the pulsar's spin-down power $\dot{E} = 1.2\times 10^{38}\;{\rm erg\;s}^{-1}$, and estimated the pulsar's 0.5 -- 8 keV X-ray efficiency $\eta_X \equiv L_{X}/\dot{E} = 5.5 \times 10^{-5} D_{10}^2$, where $D_{10}$ is the distance to the pulsar in units of 10 kpc.
In summary, A+11 characterized this distant pulsar as the most rapidly rotating non-recycled pulsar and the second most energetic Galactic pulsar known (after the Crab pulsar), but far less efficient at generating a PWN and converting the spin-down power to X-rays.

The pulsar has not been detected in the $\gamma$-rays, perhaps due to its location amidst a particularly crowded region in the $\gamma$-ray sky.
An unidentified {\sl Fermi} source 2FGL J2022.8+3843c is listed in the Second {\sl Fermi} Catalog, and given a tentative association with the SNR G079.6+01.0 \citep{2012ApJS..199...31N}.
\cite{2013ApJS..208...17A} discuss a possible pulsar counterpart $\sim 0\fdg06$ from the pulsar position, which they claim to show a PL spectrum with exponential cut-off, but still without any pulsations.

To study the pulsar's phase-resolved X-ray spectrum and further investigate its unusually faint PWN, we carried out a 110 ks {\sl XMM-Newton} observation of J2022+3842.
In this deep observation we searched for X-ray counterpart of the radio SNR, an extended PWN and the pulsar's off-pulse emission. We also performed X-ray timing of the pulsar and phase-resolved spectral analysis.

\section{Observation and Data Analysis}

Pulsar J2022+3842 was observed with the European Photon Imaging Camera (EPIC) of the {\sl XMM-Newton} observatory (obsid 0652770101) on 2011 April 14 (MJD\,55665) for about 110 ks. EPIC-pn chip \#4 and EPIC-MOS2 chip \#1 were operated in timing mode while the EPIC-MOS1 camera and the rest of the MOS2 chips were operated in imaging mode.
The EPIC data processing was done with the {\sl XMM-Newton} Science Analysis System (SAS) 12.0.0\footnote{\url{http://xmm.esac.esa.int/sas}}, applying standard tasks.

\begin{figure}[ht]
{\includegraphics[width=85mm]{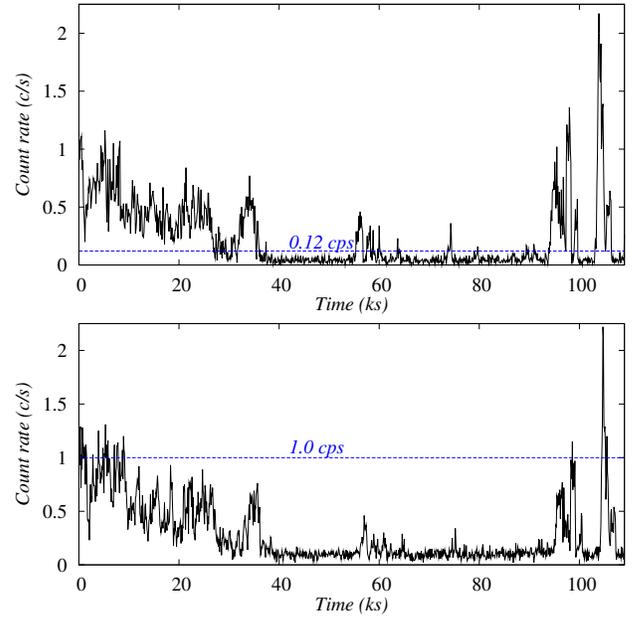}}
\caption{Flaring light curves in EPIC-pn (top) and MOS1 (bottom) for the entire observation duration.
Optimal GTI cut-off rates are shown by dashed blue lines.}
\label{fig1}
\end{figure}

The observations were partly affected by soft-proton flares.
These flaring events are characterized by periods of significantly higher background and rapid variability.
Periods of strong flaring are better identified using light curves of single pixel events (Pattern = 0) with energies $> 10$ keV, henceforth referred to as flaring light curves\noindent\footnote{\url{http://xmm.esac.esa.int/sas/current/documentation/threads/}}.
In Figure \ref{fig1}, we show the EPIC-pn (chip \#4) and MOS1 flaring light curves and the count rate cut-offs used to select Good Time Intervals (GTIs).
We simultaneously optimized the GTIs and source events extraction regions to extract the highest signal-to-noise (S/N) spectra.
Events extraction from MOS2 can accommodate more flaring intervals when a small extraction region is selected for the source, while the EPIC-pn timing mode data automatically include a large background region along the chip columns and hence require removal of most of the flaring intervals to maintain a high S/N.
The GTI-, energy- and region-filtered data have net exposures of 61.72 ks in EPIC-pn, 105.10 ks in MOS1, and 97.8 ks in MOS2.

\begin{figure}[ht]
{\includegraphics[width=85mm]{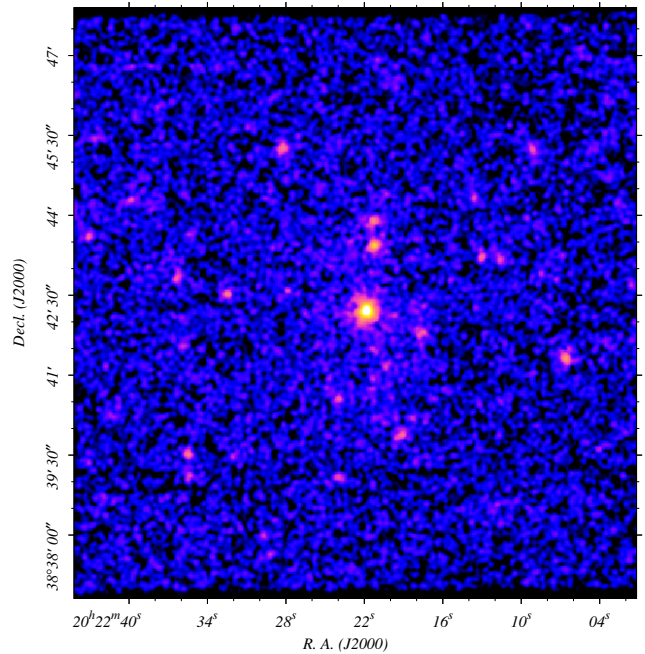}}
\caption{Binned and smoothed MOS1 image of the field around PSR J2022+3842 (center) in the 0.5--10 keV band.}
\label{M1image}
\end{figure}

Using the SAS source detection task \texttt{emldetect} on the MOS1 image (Figure \ref{M1image}), we determined the target source coordinates, $\alpha = 20^{\rm h} 22^{\rm m} 21\fs585, \delta = +38^\circ 42^\prime 14\farcs61$, with a statistical 1$\sigma$ uncertainty of $0\farcs18$.
This position differs from the CXO position by 1\farcs08, which is consistent with the {\sl XMM-Newton}'s systematic position uncertainty of $\approx 1^{\prime\prime}$ \citep{2009A&A...493..339W}.

\subsection{Timing Analysis}\label{TimingA}

\begin{figure}[ht]
{\includegraphics[width=43mm]{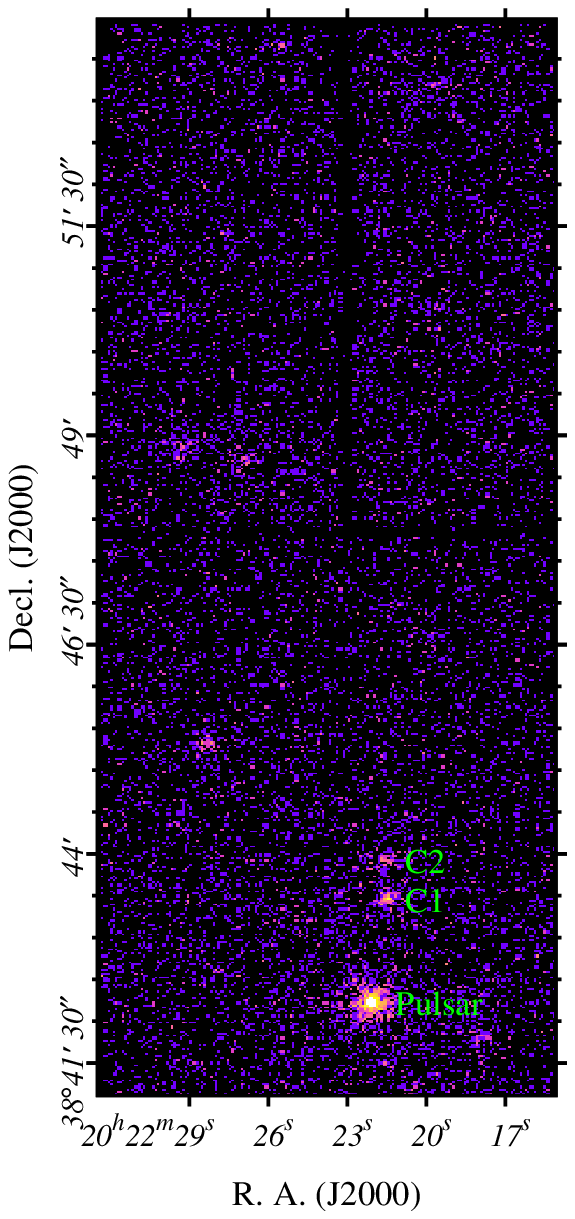}}{\includegraphics[width=43mm]{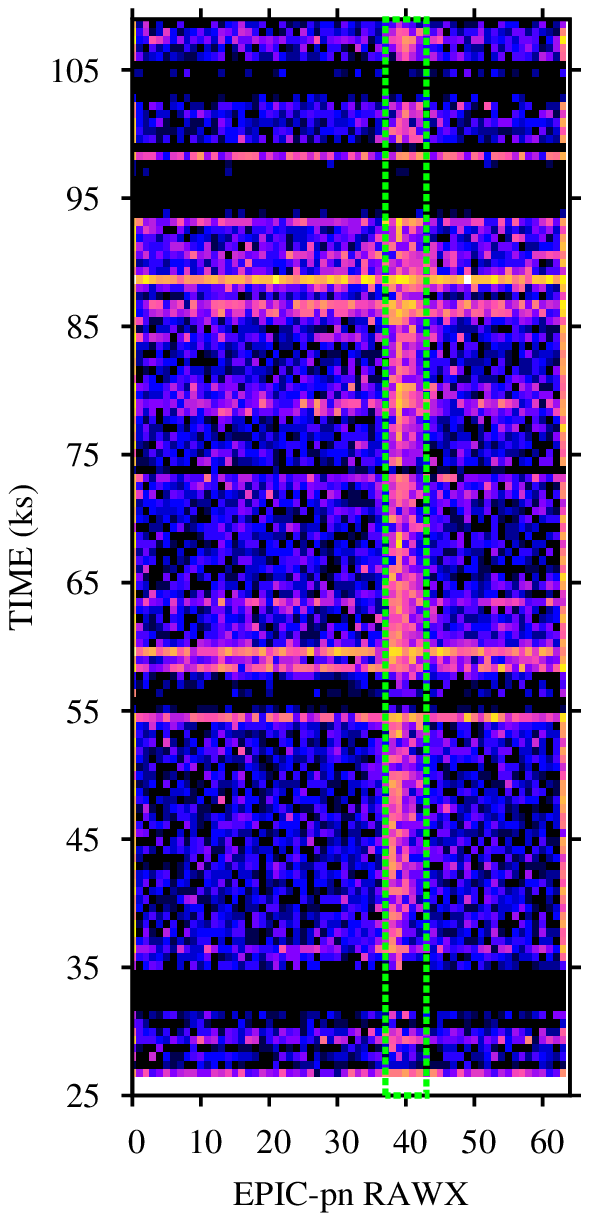}} \\
{\includegraphics[width=85mm]{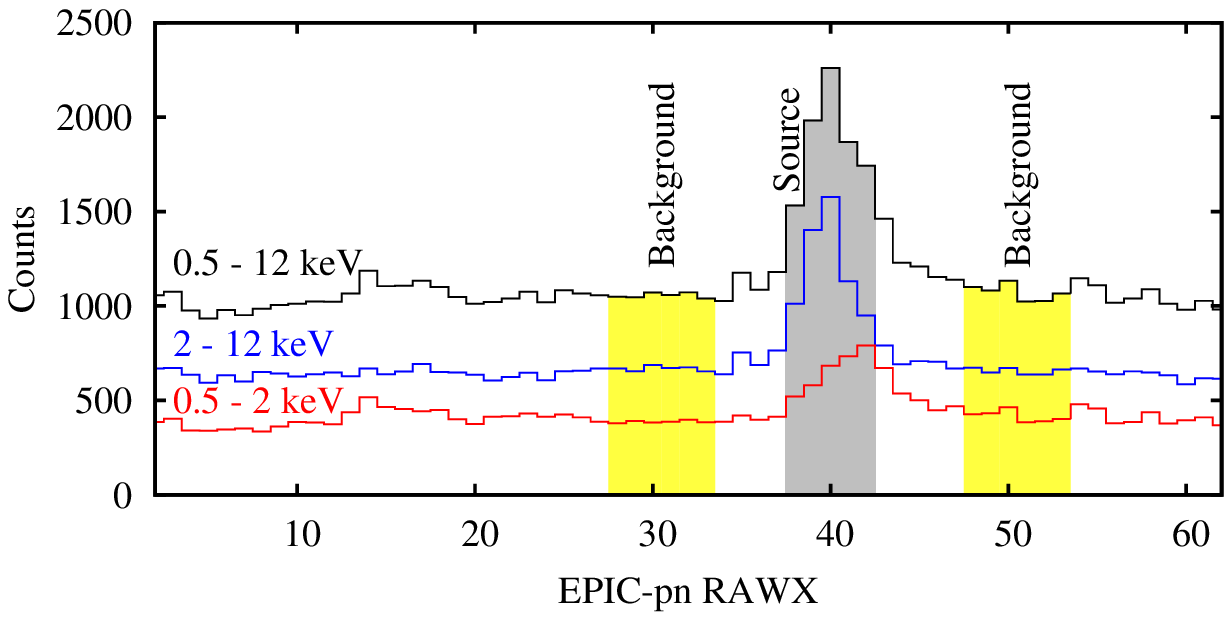}} \\
\caption{Top-left: Combined MOS1 and MOS2 (imaging mode chips) image, cropped to the PN timing-chip FOV, with the target pulsar and contaminant sources C1 and C2.
Top-right: GTI filtered, 0.5--12 keV PN timing data; time origin reset to the observation start time.
Bottom: PN events histogram along RAWX.
The shaded areas show the events extraction ranges for spectral analysis for the pulsar (RAWX = 38--42, grey) and background (RAWX = 28--33 and 48--53, yellow).}
\label{pnTiming}
\end{figure}

In the EPIC--pn (PN hereafter) timing mode, the events collected over the entire chip \#4 are collapsed into the read-out row (coordinate axis RAWX) and are read out at a high speed, providing a time resolution of 30 $\mu$s at the expense of positional information along the coordinate axis RAWY. 
In Figure \ref{pnTiming} (top-right panel), we show the GTI-filtered 0.5--12 keV PN data by plotting the events' RAWX positions against their times of arrival (TOAs).
Note that this representation is different from the conventional RAWX versus RAWY plot.
The plotted time coordinate represents elapsed time since the start of observation, and the horizontal gaps in the plot represent flaring intervals from which data has been discarded; the initial 25 ks of the filtered flaring interval is omitted from the plot (see Figure \ref{fig1}, top panel). 

Since positional information is available only along one coordinate for all events in PN, we located the target and other sources in the field by analyzing the MOS1 imaging mode data.
By identifying the PN timing-chip's field-of-view (FOV) on the MOS1 image (Figure \ref{pnTiming}, top-left panel), we found two potential contaminant sources, C1 and C2 (Figure \ref{M1image}), with the projected RAWX separations from the target of about $9\arcsec$ and $8\arcsec$ ($\approx 2$ PN pixels of $4\farcs1 \times 4\farcs1$ size).
C1 and C2 spectra are soft, with significant emission only below 2 keV, while the pulsar's spectrum is harder, with strong attenuation below 1 keV (section \ref{spectralanalysis}).
Hence, we distinguish the target and contaminant positions and contributions by plotting the PN RAWX position histograms for events with energies 0.5--12 keV, 2--12 keV and 0.5--2 keV (Figure \ref{pnTiming}, bottom panel). 
The pulsar is centered at RAWX = 40, as seen clearly in the 2--12 keV histogram, while C1 and C2 contributions peak at RAWX = 42, as seen in the 0.5--2 keV histogram.

For timing analysis, we extracted events from the RAWX segments 36--41, which excludes a significant fraction of events from the adjacent contaminant sources and provides the highest significance of pulsations.
A total of 9755 events were extracted over a time span of 82512 s, in the 0.5--12 keV range.

We applied the standard SAS task \texttt{barycen} to transform the X-ray event times from spacecraft Terrestrial Time (TT) to Barycenter Dynamical Time (TDB). 
We found the previously reported 41 Hz pulsations (A+11) using $Z_1^2$ test (e.g., \citealt{1983A&A...128..245B}).
However, subsequent phase folding over twice longer period reveals two distinct pulses with markedly unequal fluxes.
We conclude that the pulsar has a twice smaller pulsation frequency, about 20.5 Hz, with two narrow peaks per period (main pulse and interpulse). 

\begin{figure}[ht]
{\includegraphics[width=87mm]{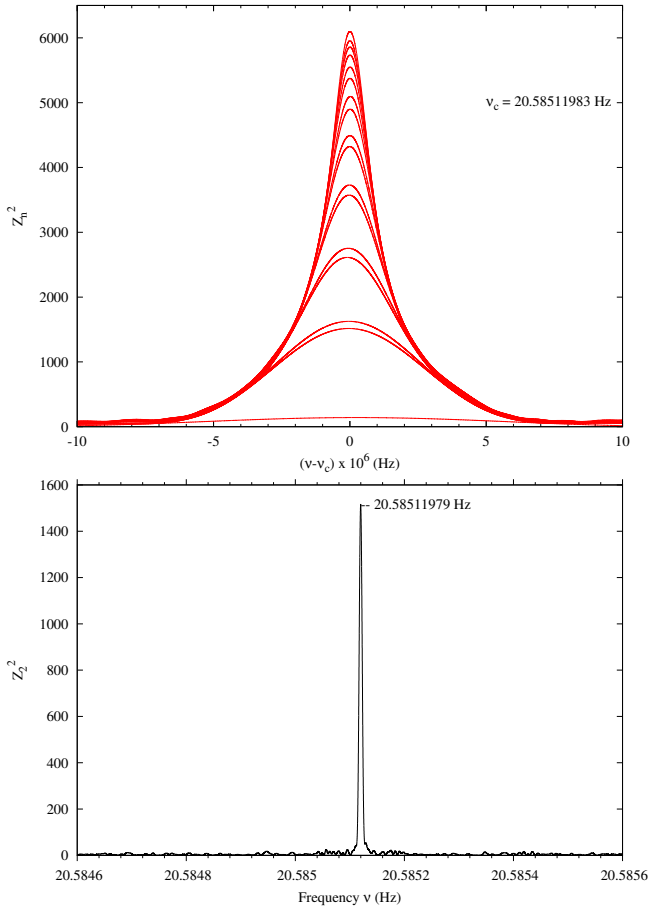}} \\
\caption{Bottom: Pulsation frequency search using the $Z_2^2$ test.
Top: The $Z_n^2$ statistics around the central frequency $\nu_c = 20.58511983$ Hz for harmonics $n = 1$--17 at $\dot\nu = -4.05 \times 10^{-11}$ Hz s$^{-1}$}.
\label{fig4}
\end{figure}

To measure the frequency more precisely and estimate the frequency derivative, we switched to $Z^2_n$ tests ($n > 1$ is the number of harmonics included), which are more sensitive in the case of narrow double peaks.
We searched the $\nu$-$\dot{\nu}$ space in the box $20.584\;{\rm Hz}< \nu < 20.586$ Hz, $-4.7 \times 10^{-11}\,{\rm Hz\;s}^{-1} < \dot\nu <- 2.7 \times 10^{-11}$ Hz s$^{-1}$, with step sizes of $1 \times 10^{-10}$ Hz and $1 \times 10^{-14}$ Hz s$^{-1}$, and found $Z^2_{2,{\rm max}} = 1515$ for $\nu = 20.58511979(9)$ Hz, $\dot\nu = -4.4(8) \times 10^{-11}$ Hz s$^{-1}$ (Figure \ref{fig4}, bottom panel) at the reference epoch 55666.23783581 (MJD TDB). 
Here and below the numbers in parentheses are $1\sigma$ uncertainties for the corresponding last significant digit(s) of the measured quantity.

We show the results of $Z_n^2$ tests, for $n = 1$--17, in the top panel of Figure \ref{fig4}.
The H test \citep{1989A&A...221..180D} fails to find a reasonable value for the number of statistically significant harmonics, as the H-statistic is an increasing function of $n$ even beyond $n=30$.
Adopting $Z_n^2$ test with multiple harmonics ($n\geq 12$), as is expected for very narrow pulse profiles, we consistently find the test statistics reaching maxima at $\nu_{\rm XMM} = 20.58511983(1)$ Hz, $\dot{\nu}_{\rm XMM} = -4.05(13) \times 10^{-11}$ Hz s$^{-1}$.

The corrected pulsar ephemeris at the {\sl RXTE} observation reference epoch of 55227.00000027, $\nu_{\rm RXTE} = 20.5865044829(71)\,{\rm Hz}, \dot\nu_{\rm RXTE} = -3.6501(79) \times 10^{-11}$ Hz s$^{-1}$, is straight-forwardly inferred from the values reported by A+11. 
From this ephemeris, the expected frequency at the reference epoch of the {\sl XMM-Newton} observation is 20.5851193(30) Hz, which coincides with the measured $\nu_{\rm XMM}$ at a $0.2\sigma$ level.
Conversely, using the frequency values at the {\sl RXTE} and {\sl XMM-Newton} epochs, we calculate the long-term frequency derivative $\dot{\nu}_{\rm XMM-RXTE}= (\nu_{\rm XMM}-\nu_{\rm RXTE})/\Delta T = -3.64861(3) \times 10^{-11}$ Hz s$^{-1}$ (where $\Delta T = 439.238$ days is the difference between the epochs).
Being more precise than $\dot\nu_{\rm RXTE}$ due to the much longer time span, this estimate is consistent with $\dot\nu_{\rm RXTE}$ at the $0.2\sigma$ level, which suggests that there were no glitches between the {\sl RXTE} and {\sl XMM-Newton} observations.
It, however, differs by about $3\sigma$ from $\dot{\nu}_{\rm XMM}$.
Given the excellent agreement between $\dot\nu_{\rm RXTE}$ and $\dot{\nu}_{\rm XMM-RXTE}$, and the relatively short time span of the {\sl XMM-Newton} observation (82 ks versus 691 ks for the {\sl RXTE} observation), we consider $\dot\nu_{\rm RXTE}$ (or $\dot{\nu}_{\rm XMM-RXTE}$ if we believe there were no glitches between the two observations) more reliable than $\dot{\nu}_{\rm XMM}$. The timing solution and derived pulsar properties are listed in Table \ref{j2022timing}.

\begin{table}[]
\setlength{\tabcolsep}{1.2em}
\caption{Timing solution and derived parameters for PSR J2022+3842.\label{j2022timing}}
\begin{tabular}{ll}
\tableline
\tableline
Parameter	& Value			\\ \tableline\tableline \\
Period $P$ (ms)	& 48.578779636(24)		\\[1.5ex]
Period derivative $\dot{P}$	& 8.61(2) $\times 10^{-14}$	\\ [1.5ex]
Epoch	(MJD TDB)	& 55666.23783581	\\ [1.5ex]
\hline \\
Main Pulse (FWHM)	& $0.020\pm0.002$ \\ [1.5ex]
Interpulse (FWHM)	& $0.023\pm0.002$ \\ [1.5ex]
Pulse separation	& $0.480\pm0.003$ \\ [1.5ex]
\hline \\
Spin-down energy rate $\dot{E}$ (erg\,s$^{-1}$)	& $3.0 \times 10^{37}$		\\ [1.5ex]
Characteristic age $\tau$ (kyr)	& 8.9	\\ [1.5ex]
Surface dipole magnetic field $B_s$ (G)	& $2.1 \times 10^{12}$ \\ [1.5ex]
\tableline
\tableline \\
\end{tabular}
\end{table}

\begin{figure}[ht]
{\includegraphics[width=90mm]{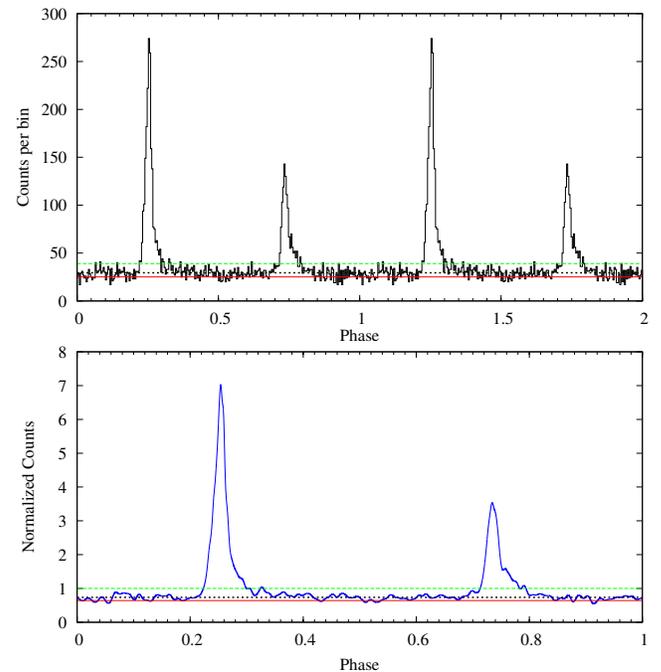}} \\
\caption{Top: Phase-folded and binned pulse profile with averaged background count rate (red line), average off-pulse count rate (black, dotted line), and count rate cut-off used to establish pulse base width (dashed, green line).
Bottom: KDE smoothed, normalized pulse profile.}
\label{pulses}
\end{figure}

The 0.5--12 keV folded ($\nu = 20.58511983$ Hz, $\dot\nu = -3.65\times 10^{-11}$ Hz s$^{-1}$, zero phase epoch = 55666.23783581) and binned (250 equal bins) X-ray pulse profile is shown in the top panel of Figure \ref{pulses}. 
In order to determine the pulse phase and pulse separation accurately, we first smooth the data using an adaptive kernel density estimation (KDE) technique \citep{2012msma.book.....F}.
We assign Gaussian kernels to each event with a bandwidth adapted to the number density of events at its phase.
The smoothed and area-normalized pulse profile is shown in the bottom panel of Figure \ref{pulses}. 
The main pulse and interpulse peak at phases $\phi_{\rm main}=0.254\pm0.001$ (FWHM = $0.020 \pm 0.002$) and $\phi_{\rm inter}=0.734\pm0.002$ (FWHM = 0.023 $\pm$ 0.002), i.e., the pulse separation is $0.480 \pm 0.003$. 
We determine the base widths of the main pulse and interpulse to be $\approx 0.074$ and $\approx 0.070$, respectively, using a count-rate cut-off (dashed, green line Figure \ref{pulses}) just above the off-pulse average of $\approx 29.4$ counts per bin (dotted black line).
We estimate the pulsed fraction  $p=0.77 \pm 0.02$, defined as the ratio of background-subtracted counts 
in the two pulses ($N_{\rm pulsed}=2703$) to the background-subtracted net source counts ($N_{\rm net}=3488$,
$N_{\rm bgd} = 6267$). The intrinsic pulsed fraction $p_{\rm int}$ of the pulsar radiation is higher because of some contribution from the unresolved PWN.
Using the PWN flux measured from the CXO ACIS data (see Section 2.2), we estimate $p_{\rm int}=0.84\pm 0.03$.
The 1$\sigma$ uncertainties for the pulse profile parameters quoted above are found through Monte-Carlo estimations with non-parametric bootstrap re-sampling of our data \citep{2012msma.book.....F}.

We performed a similar analysis of the MOS2 timing mode data.
The MOS2 CCD has a lower sensitivity than PN and a considerably lower time resolution of 1.5 ms. 
We achieved highest S/N for 2820 total counts extracted in the 1.1--8 keV range, over 109.7 ks of the observation, of which only $646\pm35$ were from the source.
The $Z^2_2$ test returned a high statistic $Z^2_{2,{\rm max}} = 862$ for $\nu = 20.58511995(8)$ Hz and $\dot\nu = -3.65(60) \times 10^{-11}$ Hz s$^{-1}$, at the reference epoch 55666.23783581, consistent with the PN timing ephemeris.
We, however, found the phase-folded pulse profile to be noisy due to low source counts, with the pulses broadened due to the poorer time resolution of MOS2.
We also found an absolute timing error of $\approx +6.8$ ms, comparing the phase shift of the MOS2 pulse profile with respect to the PN profile. 
Due to the lower S/N and the lack of recent calibration information\footnote{\url{http://xmm2.esac.esa.int/docs/documents/CAL-TN-0082.pdf}}, we exclude the MOS2 data from further analysis.

\subsection{Spectral Analysis}\label{spectralanalysis}

We use XSPEC v.12.7.1\footnote{\url{http://heasarc.gsfc.nasa.gov/docs/xanadu/xspec}} for X-ray spectral analysis.
We model absorption by the interstellar medium (ISM) using the T\"{u}bingen-Boulder model \citep{2000ApJ...542..914W} through its XPEC implementation \texttt{tbabs}, setting the abundance table to \texttt{wilm} \citep{2000ApJ...542..914W} and photoelectric cross-section table to \texttt{bcmc} \citep{1992ApJ...400..699B}, with new He cross-section based on \citet{1998ApJ...496.1044Y}. We perform chi-square fitting of the spectra (C-statistic for contaminant C2), and quote the 90\% confidence uncertainties for the model parameters evaluated for single interesting parameter.

Prior to pulsar spectral analysis, we modeled the spectra of the contaminating sources C1 and C2 (see Figure \ref{pnTiming}), using the MOS1 and archival (ObsID \#5586) CXO ACIS-S data.

\begin{figure}[ht]
{\includegraphics[width=125mm,angle=0]{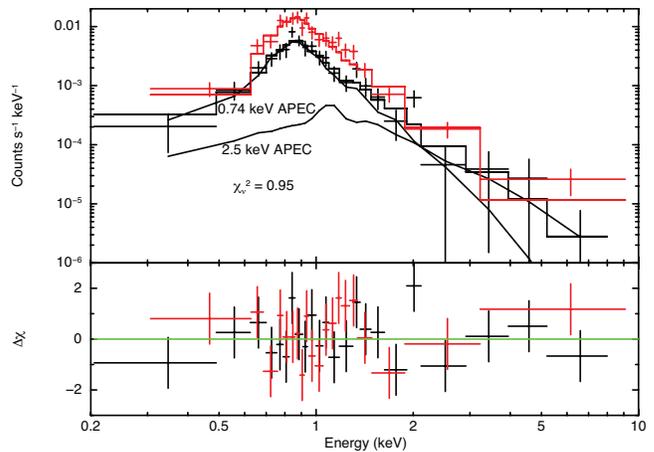}} 
\caption{ACIS-S (red) and MOS1 (black) spectra of contaminant C1 fitted with two APEC components.
Individual model components for MOS1 are shown with black curves.
The bottom panel shows residuals of the $\chi^2$ fit in units of sigma.}
\label{figC1}
\end{figure}

In the MOS1 image, contaminant C1 at coordinates $\alpha = 20^{\rm h} 22^{\rm m} 20\fs9$, $\delta = +38^{\circ} 43\arcmin 28\farcs55$ is offset by $75\arcsec$ from the pulsar, but this separation projected onto the one-dimesional (1D) PN image is just 9$\arcsec$.
For spectral fitting, we extracted events from a 12$\arcsec$ radius circle around the source in MOS1 (308 net source counts
in the 0.2--10 keV band) and from a $4\farcs2$ radius circle in ACIS-S (349 net counts in 0.3--10 keV band).
This source is coincident with HD 194094, a B0V star likely associated with the open cluster M29 at $D\approx 1.15$ kpc in Cygnus\footnote{\url{http://simbad.u-strasbg.fr/simbad/}}.
We fit the stellar spectrum with a two-component APEC model (calculated using ATOMDB code v2.0.1\footnote{\url{http://atomdb.org}}) which describes the emission from shocked, collisionally-ionized winds seen in such early-type stars.
For the best fit ($\chi^2_\nu = 0.8$ for 38 d.o.f., Figure \ref{figC1}), the two components have the temperatures $kT_1 = 0.74_{-0.05}^{+0.06}$ keV and $kT_2 = 2.6 \pm 1.4$ keV, for abundances fixed at solar values, and the absorption column density $n_H = 1.56_{-0.62}^{+0.72} \times 10^{21}$ cm$^{-2}$.
The absorbed flux is $F^{\rm abs}_{0.5-10 {\rm keV}} = 2.4^{+0.1}_{-0.2} \times 10^{-14}$ erg cm$^{-2}$ s$^{-1}$.

\begin{figure}[ht]
{\includegraphics[width=125mm,angle=0]{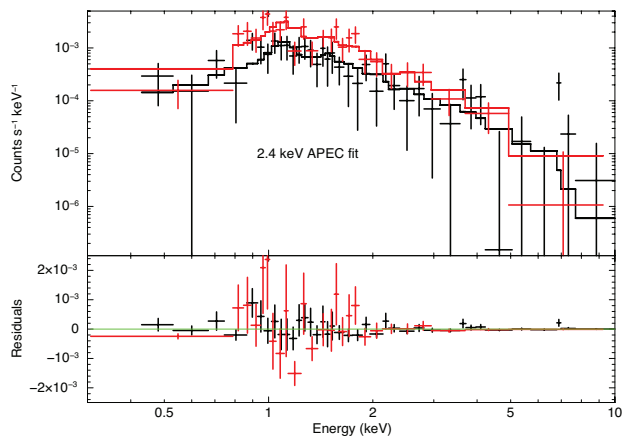}}
\caption{ACIS-S (red) and MOS1 (black) spectra of contaminant C2 fitted with APEC model.
The bottom panel shows C-statistic fit residuals.}
\label{figC2}
\end{figure}

Contaminant C2 at coordinates $\alpha = 20^{\rm h} 22^{\rm m} 20\fs9$, $\delta = +38^\circ 43\arcmin 57\farcs46$ is offset by $103\arcsec$ from the pulsar but has a projected separation of $8\arcsec$ in the 1D PN image. It is an unidentified soft X-ray point source. 
For spectral fitting, we extracted events from a $10\arcsec$ radius circle around the source in MOS1
(133 net counts in 0.4--10 keV band), and from a $3\farcs8$ radius circle in ACIS-S (136 net counts in 0.3--10 keV). 
A C-statistic \citep{1979ApJ...228..939C} fit with a single-component APEC model, with abundances fixed at solar values, yields $kT=2.4_{-0.6}^{+0.9}$ keV, $n_H = 3.2_{-1.0}^{+1.7} \times 10^{21}$ cm$^{-2}$, and $F^{\rm abs}_{0.5 - 10 {\rm keV}}$ = 1.7$^{+0.1}_{-0.1} \times 10^{-14}$ erg cm$^{-2}$ s$^{-1}$ (see Figure \ref{figC2}).

The low-energy part of the pulsar's emission ($E\lesssim 0.5$ keV) is strongly absorbed by the ISM because of the large distance and proximity to the Galactic plane ($b = 0\fdg96$). 
To fit the phase-integrated spectrum, we used the MOS1 and ACIS-S data in the 0.5--10 keV band, while for the PN data 
we chose a narrower 2--10 keV band to reduce the contamination from the soft X-ray sources C1 and C2, whose contribution is significant below 2 keV (see Figures \ref{figC1} and \ref{figC2}).
The extraction parameters and net counts for different instruments are given in Table \ref{specfilter}.

\begin{figure}[ht]
{\includegraphics[height=85mm,angle=270]{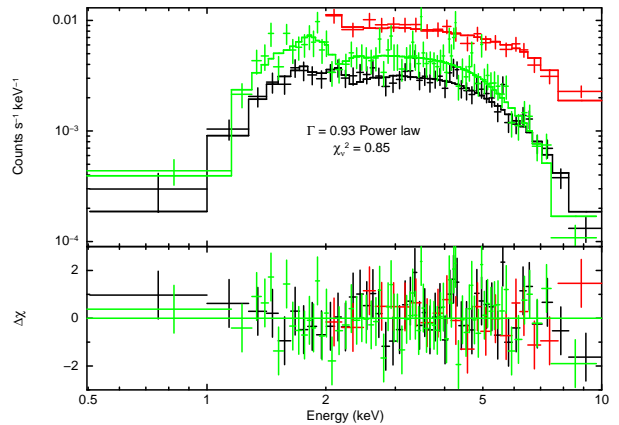}} 
\caption{Absorbed PL fit and its residuals (in units of sigma) for the phase-integrated pulsar spectra from ACIS-S (black), MOS1 (green) and PN (red).}
\label{plfit0}
\end{figure}

\begin{figure}[ht]
{\includegraphics[width=100mm]{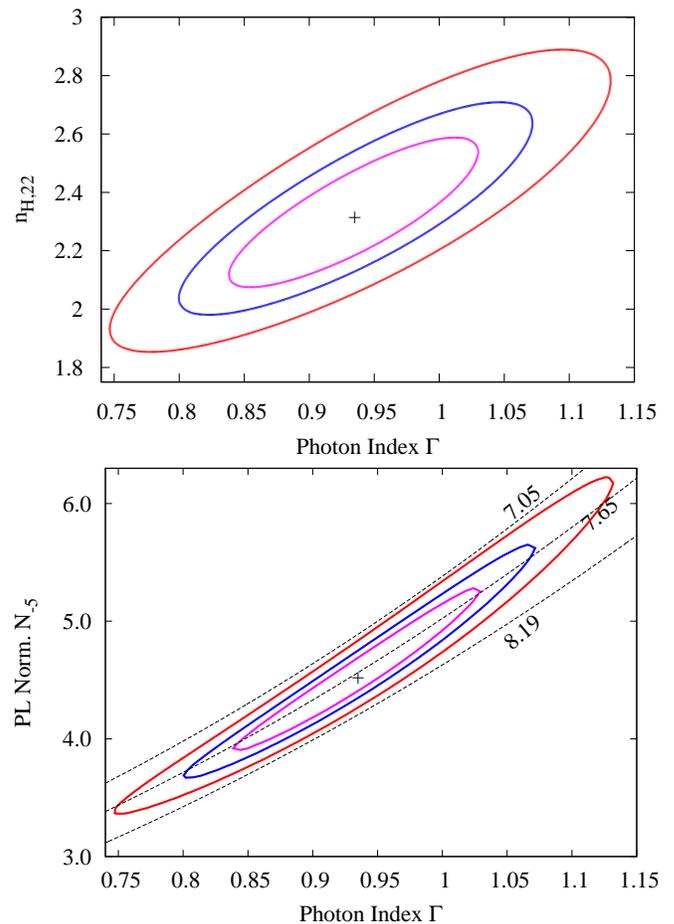}} 
\caption{Top:  $n_H$ -- $\Gamma$ confidence contours at the 68\%, 90\%, and 99\% levels for the phase-integrated spectral fit. 
Bottom: $N_{-5}$ -- $\Gamma$ confidence contours at the 68\%, 90\%, and 99\% levels for the phase-integrated spectral fit. $N_{-5}$ is the PL normalization in units of $10^{-5}$  photons cm$^{-2} {\rm s}^{-1} {\rm keV}^{-1}$ at 1 keV.}
\label{PplContout}
\end{figure}

We find that an absorbed PL model with $\Gamma= 0.93^{+0.10}_{-0.09}$ fits the phase-integrated spectra very well (Table \ref{specfit}, Figure \ref{plfit0}). 
Inclusion of the contamination-free MOS1 and ACIS-S spectra with a lower energy cut-off allowed us to constrain the
hydrogen column density, $n_{H,22} = 2.32^{+0.29}_{-0.26}$. The two parameter confidence contours for this PL fit are shown in Figure \ref{PplContout}.

The photon index we measured is consistent with that obtained by A+11 from the ACIS-S data, but the hydrogen column density is substantially different from  $n_{H,22}= 1.6\pm 0.3$ obtained by \citet{2011ApJ...739...39A}. 
Our separate fit of the ACIS-S pulsar spectrum gave all the fitting parameters close to those obtained in the PN+MOS1+ACIS-S fit, including $n_{H,22} = 2.2 \pm 0.4$.
The discrepancy in the $n_H$ values is due to the different absorption model (\texttt{phabs} with abundance table \texttt{angr}; \citealt{1989GeCoA..53..197A}) used by A+11.

\begin{figure}[ht]
{\includegraphics[height=85mm,angle=270]{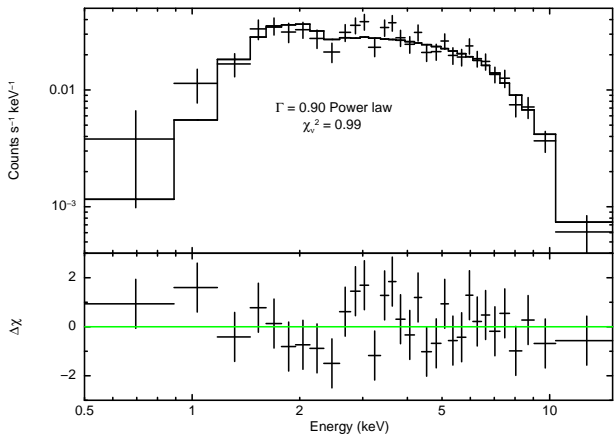}} 
\caption{Absorbed PL fit and its residuals for the  main pulse spectrum.}
\label{Pplfit0}
\end{figure}

To examine the dependence of spectral parameters on pulsation phase, we divided the pulse profile into main pulse, interpulse and off-pulse regions and analyzed their spectra individually. 
The main pulse contributes $\approx 40\%$ of the total pulsar counts in just $\sim 10\%$ phase interval. 
This allowed us to extract a high-quality (${\rm S/N} > 30$) main-pulse spectrum in the 0.5--10 keV range, from a narrow $12\farcs3$ segment around the target position in the 1D PN image, with low contamination. 
A PL model with $\Gamma \approx 0.9$ fits the main pulse spectrum well (Table \ref{specfit}, Figure \ref{Pplfit0}).

\begin{figure}[ht]
{\includegraphics[height=85mm,angle=270]{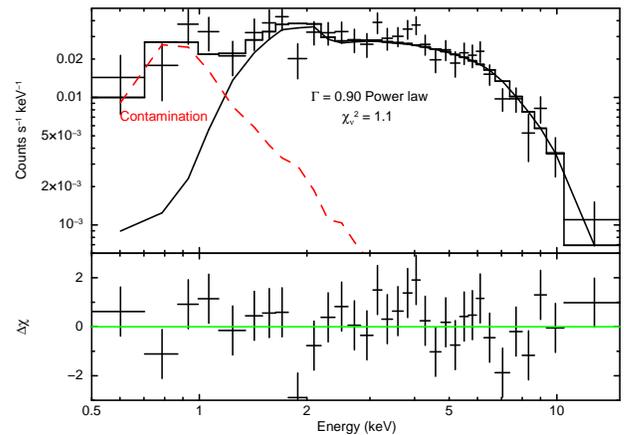}} 
\caption{Absorbed PL fit and its residuals for interpulse spectrum.
The black solid and red dashed lines show the pulsar's and combined C1+C2 contributions, respectively.}
\label{Splfit}
\end{figure}

\begin{figure}[ht]
{\includegraphics[height=85mm,angle=270]{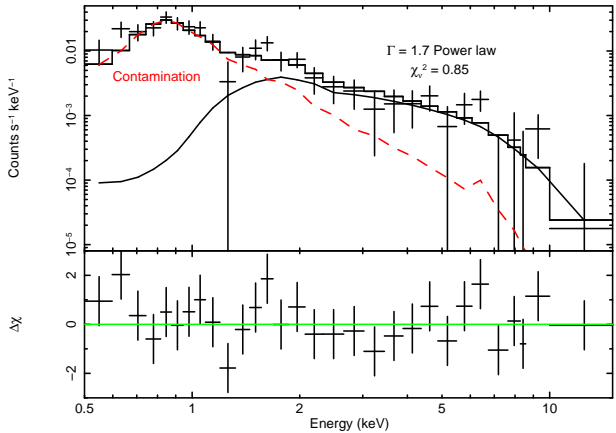}}
\caption{Absorbed PL fit and its residuals for the off-pulse spectrum.
The solid and dashed curves show the contributions from the pulsar and the C1+C2 contaminants, respectively.}
\label{Oplfit}
\end{figure}

Since the number of counts in the interpulse is lower than in the main pulse, we could not simultaneously minimize the effect of contamination and reach a sufficiently high S/N through spatial or energy filtering.
The effect of contamination is even stronger for the off-pulse emission, which barely exceeds the background level.
So, for interpulse and off-pulse spectral analysis, we extracted events in the 0.5--10 keV range, from regions spatially 
encompassing the pulsar and the contaminating sources.
Then, we added the best-fit C1 and C2 spectral models to the model for the pulsar emission and fit the combined pulsar+contaminants spectra.
We do not include any separate model for potential PWN contribution.
In addition, we fixed the $n_H$ value for the pulsar at $n_{H,22}=2.32$, obtained from the phase-integrated fit.
The procedure outlined above provides the best constraints on the photon index for the interpulse and off-pulse emission. 
The extraction parameters are listed in Table \ref{specfilter}, and the fitting parameter values are listed in Table \ref{specfit}.
The best spectral fit and residuals for interpulse emission are shown in Figure \ref{Splfit}, and for off-pulse emission in Figure \ref{Oplfit}.
The interpulse PL slope is close to that of the main pulse while for the off-pulsar emission the spectrum appears to be softer, but with a large uncertainty in its slope.

Our search for a PWN in the {\sl XMM-Newton} data did not yield positive results.
We fit the ACIS-S PWN spectrum with a PL model at fixed $n_{H,22}=2.32$ and obtained $\Gamma = 0.9\pm 0.5$, which is marginally consistent with $\Gamma = 1.4$ assumed by A+11. 
The PWN flux measured in the elliptical region with 8\farcs3 and 5\farcs1 semimajor and semiminor axes, $F_{\rm 0.5-10 keV}^{\rm abs} = 5^{+2}_{-1}\times 10^{-14}$ erg cm$^{-2}$ s$^{-1}$, is consistent with that estimated by A+11.

\begin{table*}[ht]
\setlength{\tabcolsep}{1em}
\begin{center}
\caption{Extraction parameters for phase-integrated, main pulse, interpulse and off-pulse spectra}\label{specfilter}
\begin{tabular}{@{}lccccccc@{}}\toprule
		&					\multicolumn{3}{c}{Integrated}				  &	Main pulse	  	  &	Interpulse		  & Off-pulse	 & PWN\\
\cmidrule(r){2-4}\cmidrule{5-7}
		&			MOS1 & PN	& ACIS-S			&				\multicolumn{3}{c}{PN}						 & ACIS-S\\ \midrule
\multirow{3}{*}{Phase Range$^{\rm b}$}& \multirow{3}{*}{0 -- 1}	& \multirow{3}{*}{0 -- 1}	& \multirow{3}{*}{0 -- 1} & \multirow{3}{*}{0.23 -- 0.32} & \multirow{3}{*}{0.72 -- 0.80} & 0 -- 0.22	 & \multirow{3}{*}{ -- }\\ 
		&					&				&			  &				  &				  & 0.35 -- 0.71 & \\
		&					&				&			  &				  &				  & 0.82 -- 1	 & \\[1.5ex]
Energy range (keV)&	0.5 -- 10			&	2 -- 12			&	0.5 -- 10	  &	0.5 -- 12		  &	0.5 -- 12		  & 0.5 -- 12	 & 0.5 -- 10\\[1.5ex]
Extraction region\tablenotemark{a}&	$14\arcsec$	&	38 -- 42		&	$2\farcs5$ 	  &	39 -- 41		  & 	37 -- 43		  & 37 -- 43	 & $8\farcs3 \times 5\farcs1$\\[1.5ex]
Net Counts\tablenotemark{c}	&	1606 $\pm$ 42			& 	2777 $\pm$ 91		&	1183 $\pm$ 35 	  &	1320 $\pm$ 41		  &	1130 $\pm$ 45		  & 1383 $\pm$ 109 & 96 $\pm$ 11\\[1.5ex]
\bottomrule
\end{tabular}
\end{center}
\hspace{2.5em}$^{\rm a}$ PN extraction region specified in RAWX coordinate, in pixels (1 pixel = $4\farcs1$); MOS/ACIS radius of extraction circles in arcseconds.

\hspace{2.5em}$^{\rm b}$ Pulsed to off-pulse transitional phases are omitted to obtain better constraints on fit parameters.

\hspace{2.5em}$^{\rm c}$ $1 \sigma$ uncertainties assuming Poisson statistics.
\end{table*}

\begin{table*}[t]
\setlength{\tabcolsep}{1em}
\begin{center}
\caption{Fitting parameters with 90\% confidence uncertainties for PSR J2022+3842 and its PWN.\label{specfit}}
\begin{tabular}{@{}lcccccc@{}}\toprule
Phase range & $n_{H,22}$	&	$\Gamma$	& PL. norm.\tablenotemark{a}	& $\chi^2_\nu$/d.o.f.	& $F^{\rm abs}_{0.5 - 10 {\rm keV}}$\tablenotemark{b}	&	$F^{\rm unabs}_{0.5 - 10 {\rm keV}}$\tablenotemark{b}	\\
\midrule
Integrated (100\%)\tablenotemark{c}	&	$2.32^{+0.29}_{-0.26}$	&	$0.93^{+0.10}_{-0.09}$	&	$4.53^{+0.84}_{-0.68}$	&	0.85/138	&	$6.17^{+0.25}_{-0.25}$	&	$7.62^{+0.27}_{-0.26}$	\\[1.5ex]
Main pulse (9\%)	&	$2.23^{+0.67}_{-0.56}$	&	$0.90^{+0.19}_{-0.17}$	&	$17.7^{+6.8}_{-4.7}$	&	0.99/29		&	$25.7^{+1.6}_{-1.6}$	&	$31.4^{+2.5}_{-2.2}$	\\[1.5ex]
Interpulse (8\%)	&	2.32 (fixed)		&	$0.90^{+0.13}_{-0.14}$	&	$12.8^{+2.7}_{-2.4}$	&	1.10/31		&	$18.5^{+1.5}_{-1.5}$	&	$22.7^{+1.6}_{-1.6}$	\\[1.5ex]
Off-pulse (76\%)	&	2.32 (fixed)		&	$1.70^{+0.76}_{-0.71}$	&	$2.29^{+2.55}_{-1.41}$	&	0.85/28		&	$0.89^{+0.38}_{-0.34}$	&	$1.44^{+0.42}_{-0.40}$	\\[2.0ex]
PWN			&	2.32 (fixed)		&	$0.94^{+0.46}_{-0.48}$	&	$0.36^{+0.26}_{-0.16}$	&	11.21\tablenotemark{d}/6		&	$0.48^{+0.18}_{-0.13}$	&	$0.59^{+0.17}_{-0.13}$	\\
\bottomrule
\end{tabular}
\end{center}
\hspace{4em}$^{\rm a}$ PL normalization in units of $10^{-5}$  photons cm$^{-2}$ s$^{-1}$ keV$^{-1}$ at 1 keV.

\hspace{4em}$^{\rm b}\;F^{\rm abs}_{0.5 - 10 {\rm keV}}$ and $F^{\rm unabs}_{0.5 - 10 {\rm keV}}$ are absorbed and unabsorbed fluxes, respectively, in units of $10^{-13}$ erg cm$^{-2}$ s$^{-1}$.

\hspace{4em}$^{\rm c}$ Percentages in parentheses denote the fraction of total period included.
 
\hspace{4em}$^{\rm d}$ Best-fit value of C-statistic.
\end{table*}

\section{Summary and Discussion}

We did not detect any prominent extended emission in the 105 ks MOS1 exposure of the region around PSR\,J2022+3842. 
The presence of a large number of X-ray point sources around the pulsar hinders quantitative spatial analysis for assigning restrictive upper limits on the extended emission from either the SNR or the PWN. 

Our timing analysis has shown that the true pulsar period, $P\approx 48.6$ ms, and period derivative, $\dot{P}\approx 8.6\times 10^{-14}$, are twice larger than those reported by A+11, and the phase-folded light curve has two peaks per period, the main pulse and the interpulse, separated by $\approx 0.48$ of the period.
Using these $P$ and $\dot{P}$ values, we re-evaluated the pulsar's spin-down power, $\dot{E} = 3.0\times 10^{37}$ erg s$^{-1}$, and magnetic field strength $B = 2.1\times 10^{12}$ G. 

The X-ray pulses are very narrow compared to most of the pulsars with known X-ray pulse profiles.
However, a young ($\tau=5.4$ kyr), rapidly rotating ($P=65.7$ ms) PSR J0205+6449 shows a similar X-ray pulse profile and spectral characteristics \citep{2010A&A...515A..34K}.
The very narrow X-ray pulse profiles and hard X-ray spectra of these pulsars indicate that the X-ray emission originates from the pulsar magnetosphere.
The double-peaked profile, with separations of $\approx0.48$ and no discernible bridge emission, indicate emission from diametrically opposite sites in the pulsar magnetosphere.
Gamma-ray light curves possessing similar characteristics favor a high magnetic obliquity (large angle $\alpha$ between the rotation and magnetic axes) for the pulsar \citep{2009ApJ...695.1289W}. From radio and $\gamma$-ray light curve modeling of PSR J0205+6449, \cite{2014arXiv1403.3849P} estimate $\alpha \approx 80\degr$ for the pulsar. If the similarities to PSR J0205+6449 do extend to the $\gamma$-ray regime, PSR J2022+3842 could be established as a nearly orthogonal rotator. This can be further tested through the $\gamma$-ray light curve modeling (if $\gamma$-ray emission is detected in future), or through the radio polarization measurements.

Our estimate of the total hydrogen column density (neutral, ionized and molecular) towards PSR J2022+3842, $n_{H,22} = 2.32^{+0.29}_{-0.26}$, obtained using the \texttt{tbabs} model with \texttt{wilm} elemental abundances, is significantly higher than the previous estimate, $n_{H,22} = 1.6\pm0.3$ (A+11), obtained using the \texttt{phabs} model with  \texttt{angr} abundances. 
We conclude that estimating hydrogen column densities through X-ray spectral modeling of emission from heavily obscured 
targets is highly sensitive to the ISM absorption model and abundance table used.

The phase-integrated pulsar spectrum fits a hard PL model with $\Gamma = 0.9\pm0.1$. The main pulse and the interpulse contribute $\sim 80\%$ of the total emission.
The off-pulse spectrum is poorly constrained due to contamination and an inherently weak signal. A possible source of the off-pulse emission could be the compact PWN, which cannot be resolved by {\sl XMM-Newton} because of its broad point spread function.
Comparing the PN off-pulse spectrum with the ACIS-S PWN spectrum (see Table \ref{specfit}), we find different best-fit values of photon index and flux, but the uncertainties are too large to claim the distinction between the two spectra to be statistically significant.
From re-analysis of the ACIS-S data, we also found the PWN spectrum to be harder than previously assumed, with $\Gamma = 0.94^{+0.46}_{-0.48}$. 
This result is consistent with the empirical correlation between the PWN photon index and its 2--10 keV luminosity 
(and, more tightly, the PWN X-ray efficiency (see Figure 1 and Figure 7 in \citealp{2008ApJ...682.1166L}). 

\begin{figure*}[ht]
{\includegraphics[width=170mm]{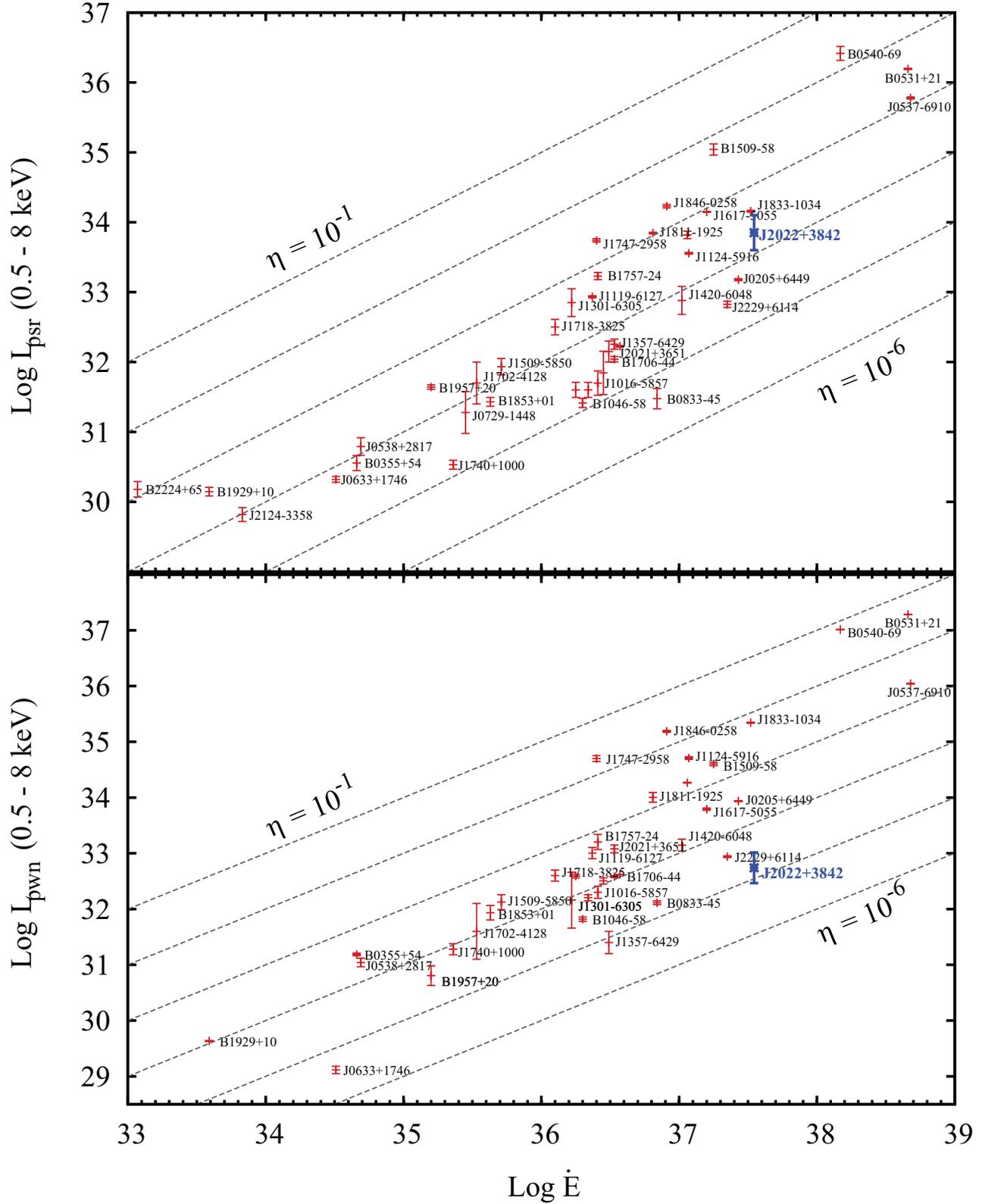}} \\
\caption{Comparison of pulsar and PWN efficiencies for non-recycled pulsars with spin-down power in the $10^{33} - 10^{39}$ erg s$^{-1}$ range.
The dashed straight lines are lines of constant efficiency. PSR J2022+3842 and its PWN are marked by blue asterisks.
These graphs are adapted from \cite{2008AIPC..983..171K} (Tables 1 and 2, and Figure 5).}
\label{psrcompare}
\end{figure*}

We have assessed that the pulsar has a factor of 4 lower spin-down power and a slightly higher X-ray flux than reported by A+11.
As a result, our pulsar X-ray efficiency estimate is a factor of 4 higher, $\eta^{\rm PSR}_{\rm 0.5-8 keV} = L^{\rm PSR}_{\rm 0.5-8 keV}/\dot{E} = 2.0 \times 10^{-4} D_{10}^2 $. 
As shown in Figure \ref{psrcompare} (top panel), the X-ray efficiency of PSR J2022+3842 is comparable to those of other young, energetic pulsars for the adopted distance of 10 kpc (for illustrative purposes, we assign 25\% uncertainty to J2022+3842's distance).
In contrast, the associated PWN efficiency, $\eta^{\rm PWN}_{\rm 0.5-8 keV}\sim 2\times 10^{-5} D_{10}^2$, is the lowest among young pulsars with comparable values of  $\dot{E}$ (Figure \ref{psrcompare}, bottom panel).
A low magnetic obliquity might in principle explain a weak PWN, but is disfavored by the observed X-ray light curve.
The reason for so low PWN efficiency remains to be understood.

\acknowledgments
We thank Michael Freyberg and Bettina Posselt for discussions and clarifications regarding EPIC-MOS2 timing analysis. We also thank Eric Feigelson for valuable discussion and suggestion on statistical techniques, and the referee for useful comments.
This work was partly supported by NASA Astrophysics Data Analysis Program award NNX13AF21G.

\clearpage

\bibliographystyle{apj}
\bibliography{J2022bibliography}

\end{document}